# DABS-LS: Deep Atlas-Based Segmentation Using Regional Level Set Self-Supervision

Hannah G. Mason[a] and Jack H. Noble[a,b]
[a]Dept. of Computer Science, Vanderbilt University
[b]Dept. of Electrical and Computer Engineering, Vanderbilt University

*Abstract*— Cochlear implants (CIs) are neural prosthetics used to treat patients with severe-to-profound hearing loss. Patient-specific modeling of CI stimulation of the auditory nerve fiber (ANFs) can help audiologists improve the CI programming. These models require localization of the ANFs relative to surrounding anatomy and the CI. Localization is challenging because the ANFs are so small they are not directly visible in clinical imaging. In this work, we hypothesize the position of the ANFs can be accurately inferred from the location of the internal auditory canal (IAC), which has high contrast in CT, since the ANFs pass through this canal between the cochlea and the brain. Inspired by VoxelMorph, in this paper we propose a deep atlas-based IAC segmentation network. We create a single atlas in which the IAC and ANFs are pre-localized. Our network is trained to produce deformation fields (DFs) mapping coordinates from the atlas to new target volumes and that accurately segment the IAC. We hypothesize that DFs that accurately segment the IAC in target images will also facilitate accurate atlas-based localization of the ANFs. As opposed to VoxelMorph, which aims to produce DFs that accurately register the entire volume, our novel contribution is an entirely self-supervised training scheme that aims to produce DFs that accurately segment the target structure. This self-supervision is facilitated using a regional level set (LS) inspired loss function. We call our method Deep Atlas Based Segmentation using Level Sets (DABS-LS). Results show that DABS-LS outperforms VoxelMorph for IAC segmentation. Tests with publicly available datasets for trachea and kidney segmentation also show significant improvement in segmentation accuracy, demonstrating the generalizability of the method.

## I. INTRODUCTION

COCHLEAR implants (CIs) are neural prosthetics used to treat patients with severe-to-profound hearing loss. The CI uses an array of 12-22 electrodes implanted into the patient's cochlea stimulate the auditory nerve fibers (ANFs) to induce the sensation of hearing. An external processor is worn by the patient and that processes sounds detected by a microphone and sends stimuli to the electrodes. After implantation, an audiologist optimizes the CI programming for a specific patient by adjusting various parameters. Typically, this process is entirely based on patient feedback and can require dozens of programming sessions, taking months or years, and often does not lead to optimal results.

We are now developing comprehensive patient-specific ANF stimulation models [1] [2], which can assist audiologists with optimizing CI programming. These models are constructed by estimating the electric potentials across the ANFs induced by the CI electrodes, which allows for explicit modeling of the electro-neural interface (ENI) and the ability to assess neural health [1]. However, this process requires accurate localization of the ANFs relative to surrounding anatomy and the CI. Localizing the ANFs is challenging because they are so small that they are not directly visible in clinical imaging.

In this work, we hypothesize the position of the ANFs can be accurately inferred from the location of the internal auditory canal (IAC), which has high contrast in CT, since the ANFs pass through this canal between the cochlea and the brain . Inspired by VoxelMorph, we propose a deep atlas-based IAC segmentation network. We create a single atlas in which the IAC and ANFs are pre-localized. Our network is trained to produce deformation fields (DFs) mapping coordinates from the atlas to new target volumes and that accurately segment the IAC. We hypothesize that DFs that accurately segment the IAC in target images will also facilitate accurate atlas-based localization of the ANFs. As opposed to VoxelMorph, which aims to produce DFs that accurately register the entire volume, our novel contribution is an entirely self-supervised training scheme that aims to produce DFs that accurately segment the target structure. This self-supervision is facilitated using a regional level set inspired loss function. We call our method Deep Atlas Based Segmentation using Level Sets (DABS-LS). In the remainder of this section, we present a discussion of related work.

### A. Deep Atlas-Based Segmentation

Atlas-based image segmentation is a process in which an "atlas" image, which has pre-defined ground truth annotations of target anatomy, is automatically registered to a novel target image. Through this registration transformation, annotations on the atlas can be transferred to the given target image. The accuracy of the resulting annotations is directly related to the accuracy of the registration transformation. Atlas-based

This work was supported in part by NIH grant R01DC014037. The work is solely the responsibility of the authors and does not necessarily reflect the views of this institute.
Hannah G. Mason is with the Department of Computer Science, Vanderbilt University, TN 37235 USA (e-mail: hannah.g.mason@vanderbilt.edu)
Jack H. Noble is with the Department of Computer Science and the Department of Electrical and Computer Engineering, Vanderbilt University, TN 37235 USA (e-mail: jack.noble@vanderbilt.edu)
This work is a substantially extended study based on an initial conference presentation at the 2023 SPIE Conference on Medical Imaging [26].

segmentation is especially common in medical image analysis, due to the similar relative positioning of many anatomical structures in humans.

Neural network architectures for image deformation are largely based on architectures for semantic image segmentation. The fully convolutional network, created by Long et al, is the basis for the most popular image segmentation network architectures [3]. This architecture uses a series of down-sampling encoders and up-sampling decoders to classify each pixel in the image, maintaining the original resolution. The encoders produce feature channels which feed both into the encoder below it and the decoder on its same resolution level. In this way, the network is able to produce high quality segmentations.

Segmentation neural networks are usually trained with images as inputs and a loss function that compares outputs to corresponding accurate ground truth labels. These networks require large amounts of annotated data to perform well. In Long et al, the dataset sizes range from about 1000-2500 images [3]. Krizhevsky et al, a breakthrough paper on deep learning in 2012, trained an impressively accurate neural network using 1.2 million images [4].

Unfortunately, accurately annotated datasets of this size are very difficult to obtain. Medical images, especially, are more difficult to capture and annotate. Ronneberger et al proposed a modification to the fully convolutional neural network to help reduce the number of training images needed [5]. In this new architecture, dubbed U-Net, the number of feature channels produced by encoders is increased greatly. This modification, combined with data augmentation, allowed the U-Net to provide very good results with only 30-35 training images.

While Ronneberger et al's initial network was for 2D images, a 3D version was published by Çiçek et al [6]. The U-Net architecture has become highly popular for segmenting 3D images [7] [8] [9].

Ideally, neural networks are trained on a sizeable dataset with accurate ground truth annotations. Sometimes a sizeable set of images is available without strong annotations. In these cases, weak- or self- supervision learning methods have been proposed. Papandreou et al. proposed an expectation-maximization method for weakly supervised training [10]. This method uses weak annotations, such as bounding boxes or image-labels, and combines the use of weakly and strongly labeled images to train the network. Self-supervision strategies utilize loss function terms designed to evaluate the quality of network predictions to learn with unlabelled data.

U-Net-based architectures for semantic segmentation have been adapted for self-supervised atlas-based segmentation. A popular example is VoxelMorph, a framework proposed by Balakrishnan et al [11]. The VoxelMorph architecture is based on the popular U-Net architecture [5], but includes a modified final convolutional layer to produce a three-channel deformation field rather than a single channel binary mask. VoxelMorph was proposed for both pair-wise registration and atlas-based segmentation. While VoxelMorph was developed for brain MRI, subsequent papers have proposed using this framework for other modalities and anatomical objects of interest. For example, Miyake et al used VoxelMorph to analyze the heart and lungs in Thoracic MDCT images [12].

In the current work, we propose a novel extension of VoxelMorph for atlas-based segmentation. The primary contribution is a custom loss function that facilitates more effective self-supervision learning for atlas-based segmentation of the particular class of structures that, when segmented accurately, are reasonably well represented locally by a Chan-Vese level set-inspired energy functional.

### B. Contributions

The loss function originally proposed for self-supervised training of the VoxelMorph architecture consists of two terms – one to reward the similarity between the deformed and target image and another to reward the smoothness of the predicted deformation field [11].

Both mean squared error (MSE) and cross correlation (CC) are common image similarity metrics. MSE measures the mean-squared differences in intensity for every pixel in the images. MSE is minimized when the intensities between two registered images are identical. In contrast, cross correlation is maximized when there exists a linear intensity relationship between two registered images.

The magnitude of the gradient of the deformation field has been proposed to measure its smoothness, and a loss function was proposed to minimize the mean magnitude of the gradient across the deformation field domain. This rewards the network for predicting smoothly varying deformation fields [11].

Some papers have been published which add to, or modify, these loss terms. Zhu et al. proposed modified loss functions for the VoxelMorph framework, where the smoothing term is based on the Laplacian of the deformation field rather than the magnitude of the gradient of the deformation field [13].

Herein, we propose a new self-supervision term for atlas-based segmentation training that is inspired by the energy functional proposed by Chan and Vese for region-based level set segmentation [14] [15]. A similarly inspired loss function was been proposed by Kim and Ye as well as our group for deep learning segmentation networks [16] [9], however, to the best of our knowledge this approach has not been applied to networks trained for atlas-based segmentation. In the remainder of the paper, we present our method and experiments designed to show its impact when applied to atlas-based segmentation of our IAC dataset as well as several other anatomical structures for which publicly available evaluation datasets were available.

## II. METHODS

### A. Datasets

We used three datasets for this paper: 1) An Internal Auditory Canal (IAC) dataset; 2) the Segmentation of Thoracic Organs at Risk in CT images dataset (SegTHOR) dataset [17]; and 3) the 2021 Kidney and Kidney Tumor Segmentation Challenge (KiTS21) [18]. The appearance of these structures well matches the Chan and Vese energy functional we propose as the region interior to the structures is relatively homogeneous in intensity compared to background structures. Thus, an accurate segmentation should overlap with an image region with minimized intensity variance.

*1) IAC.*

This dataset contains an atlas CT image, with a ground truth localization. It contains 700 patient CT images, where affine registrations to the atlas have been computed using a mutual information-based method as reported in previous works [19] [20] [21]. The IAC dataset was split into 688 images for training, 7 for validation, and 13 for testing. Ground truth binary segmentations of the IAC were created for the 20 patients in the validation and test sets.

*2) SegTHOR*

The SegTHOR dataset is part of the SegTHOR 2019 Challenge. This dataset provides 40 images with accurate ground truth segmentations for the trachea. One image was set aside as an atlas. The remaining 39 images were affinely registered to it using a mutual information-based approach [19]. Due to the small size of this dataset, data augmentation was performed. Five images from the original dataset were set aside to serve as a test set and were not included in the data augmentation. The remaining images were augmented by applying a randomly generated deformation field. This process was done 3 times for each image. The resulting dataset was split into 109 images for training and 27 images for validation.

*3) Healthy KiTS21*

Our final dataset comes from the 2021 Kidney and Kidney Tumor Segmentation Challenge. We chose not to focus on tumor segmentation at this time. Tumors are highly variable in shape, size, and position, making them unsuitable for localization with an atlas-based method. However, we selected 130 images of healthy kidneys to use to evaluate our method for atlas-based kidney segmentation. One image was chosen as the atlas, and the remaining 129 images were affinely registered to it using a mutual information-based approach [19]. The Healthy KiTS21 dataset was split into 105 images for training, 12 images for validation, and 12 images for testing.

*4) Data Preprocessing*

For all datasets, non-atlas images were affinely normalized with the atlas and then cropped to a 64 x 64 x 64 resolution cube around the centroid of the atlas segmentation. Voxel sizes were chosen to ensure the full object was visible in all images, resulting in 0.30 mm per voxel for the IAC dataset and 2.84375 mm per voxel for the others. Some pixels in the atlas-aligned images were outside the boundaries of the source images – in these cases, these pixels were assigned the mean value of the existing pixels.

In the VoxelMorph paper, it is recommended that the intensity values of the input data be normalized between 0 and 1. CT images are prone to having image artifacts, especially from metal, that can greatly change the intensity range of an image and impact normalization quality. In order to better standardize the datasets, a dataset-specific normalization was performed. For each dataset, constants $i_{min}$ and $i_{max}$ where determined representing the average minimum and maximum intensities across the training dataset. Then each training image $T$ is normalized such that $T=(T-i_{min})/(i_{max}-i_{min})$ as a preprocessing step. Thus, most of the intensity values in the resulting images fall between 0 and 1, but outlier intensities exist outside that range.

### B. Network Architecture

We implemented the U-Net architecture adopted for VoxelMorph and inspired by Ronneberger *et al.* and Wolny *et al.* [11] [7] [5]. The architecture diagram can be seen in Figure 1. The input to our network is a single channel CT image $T$. The output is a non-rigid deformation field $\phi$ which maps coordinates from the atlas space to the target image space. There are 4 levels of resolution, with three encoder levels and three decoder levels. There is a convolutional layer at the beginning of the network to extract feature channels from the input image. The output of the final convolutional layer has been modified from the single channel probability map proposed with the original U-Net to instead output a 3 channel deformation field.

### C. Loss Functions

Similarly to VoxelMorph, we adopt a cross correlation (CC)-

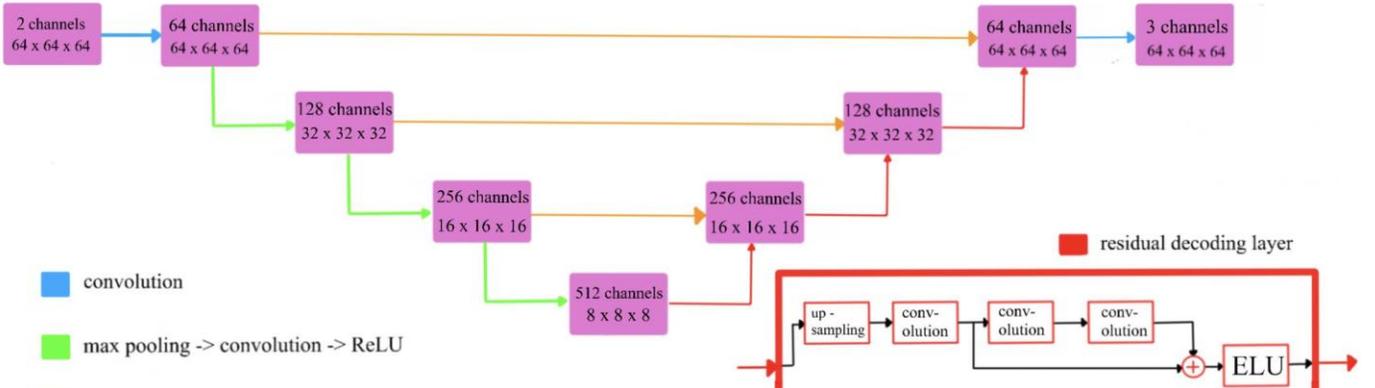

*Figure 1. Network Diagram*

based image similarity metric [11]. The equation for this term can be seen below in Equation 1, where $\mu$ denotes the mean.

$$loss_{cc} = 0.5 - \frac{\left(\phi^{-1}(T) - \mu_{\phi^{-1}(T)}\right)^T (A - \mu_A)}{2||\phi^{-1}(T) - \mu_{\phi^{-1}(T)}||\ ||A - \mu_A||} \quad (1)$$

For improved regularization, we propose a modification to the smoothness term used with VoxelMorph [11].

The smoothness term used with the VoxelMorph framework is simply the mean-squared magnitude of the gradient of the deformation field, which can be seen in Equation 2.

$$loss_{grad} = \frac{1}{N}\sum_{i}^{N}||\nabla\phi_i||^2 \quad (2)$$

However, it assigns equal weight to all areas of the image domain. For our purpose, atlas-based segmentation of a single structure, we propose modifying this term to prioritize smoothness near the target structure.

To enable measuring proximity to the target, we compute an approximate signed Euclidean distance map $D$ for the atlas segmentation $S_A$ after it has been affinely registered to the target image $T$ using fast marching-based methods. A weighting term leveraging this distance map is defined in Eqn. (3).

$$W = 0.5 + \frac{t_U - \max(t_L, \min(t_u, |D|))}{2(t_U - t_L)} \quad (3)$$

As shown in the equation, the weight is a function of the magnitude of the distance map, with values clamped between a lower threshold $t_L$ and an upper threshold $t_U$. We chose these values to be 1 and 4 mm for our datasets. This result is used as a weighting for the deformation smoothness term that more heavily weights smoothness around the border of the atlas segmentation, as shown in the loss term in Eqn. 4.

$$loss_{wgrad} = \frac{1}{3N}\sum_{i}^{N}||W_i \nabla\phi_i||^2 \quad (4)$$

Additionally, inspired by Kim and Ye, we propose a new level set-based loss term [16]. This term assumes that the interior of the desired segmentation has a distinguishable average image intensity compared to the average image intensity of the exterior region. Equations 5 through 9 define this term. It uses the atlas segmentation mask $A$ to delineate the foreground of the atlas-based segmentation on the deformed target image $\phi^{-1}(T)$. The final loss term in Eqn. (9) aims to minimize the intra-class variance of the intensity distributions of the image regions labelled as belonging to foreground and background by the atlas-based segmentation.

$$\mu_{int} = \frac{\sum_{i=1}^{N}\phi^{-1}(T)_i * A_i}{\sum_{i=1}^{N}A_i} \quad (5)$$

$$\mu_{ext} = \frac{\sum_{i=1}^{N}\phi^{-1}(T)_i * (1 - A_i)}{\sum_{i=1}^{N}(1 - A_i)} \quad (6)$$

$$\sigma_{int}^2 = \frac{\sum_{i=1}^{N}(\phi^{-1}(T)_i - \mu_{int})^2 * A_i}{\sum_{i=1}^{N}A_i} \quad (7)$$

$$\sigma_{ext}^2 = \frac{\sum_{i=1}^{N}(\phi^{-1}(T)_i - \mu_{ext})^2 * (1 - A_i)}{\sum_{i=1}^{N}(1 - A_i)} \quad (8)$$

$$loss_{LS} = \sigma_{int}^2 + \sigma_{ext}^2 \quad (9)$$

*1) Combined Loss Functions*

The loss function used in the VoxelMorph framework can be seen in Equation 14. Our proposed loss function can be seen in Equation 15.

$$loss_{VXM} = \omega_{cc} * loss_{cc} + \omega_{grad} * loss_{grad} \quad (14)$$

$$loss_{new} = \omega_{cc} * loss_{cc} + \omega_{wgrad} * loss_{wgrad} + \omega_{LS} * loss_{LS} \quad (15)$$

The weights for each of these terms were found through a heuristic optimization process which involved a sweep of various weight combinations for each dataset. The final chosen weights can be seen in Table 1. The cross-correlation weight for both datasets was kept constant at 1.0, and the rest of the weights were changed relative to it. The weights were optimized for each dataset individually. Interestingly, the weights reported in Table 1 were found to be optimal independently for all three datasets.

| Weight | Value |
|---|---|
| $\omega_{cc}$ | 1.0 |
| $\omega_{grad}$ | 1.0 |
| $\omega_{wgrad}$ | 1.0 |
| $\omega_{LS}$ | 0.5 |

## III. EVALUATION METRICS

The output deformation field from the network can be used to project the atlas IAC segmentation surface onto the target volume. We compare this resulting segmentation to ground truth segmentations using the 95th Percentile Hausdorff distance to quantify surface distance errors and Dice similarity coefficients to quantify volumetric overlap errors. The statistical significance of our results was evaluated with the dependent t-test for paired samples.

### A. Special Considerations for the IAC

The IAC is a structure with no true anatomical boundary separating it centrally from the brain. Any such boundaries delineated in the ground truth segmentation are completely arbitrary. Comparing the arbitrary boundary of the network result to the arbitrary boundary of the ground truth would not be informative. Conversely, including this area in our metrics can occlude the true performance of our results in the areas which are actually important for our downstream task of ANF localization. Thus, for the IAC, the region not important for our ANF localization was hand annotated in our validation and testing datasets, and this region is excluded when computing the evaluation metrics.

For the SegTHOR and Healthy KiTS21 datasets, the entire image is included in the evaluation metrics.

### B. Training strategy

The choice of initial weights of a convolutional neural network prior to training can have a significant impact on final network performance, and even repeat optimizations starting from identical initial weights can lead to variable results [22] [23] [24]. Henderson *et al.* discuss how this lack of consistency can make it difficult to interpret results [25]. In order to mitigate these concerns, in our experiments, pseudo-random number generators for all code libraries were seeded with a constant at the start of each trial. In addition, each training configuration was evaluated 5 times. In the results section, we report median results across training runs in order to determine the best configuration.

## IV. RESULTS

After performing a hyperparameter search for the terms in our proposed loss function as described in II.C.1, we performed an ablation study to evaluate the contribution of each of our proposed loss terms on the validation data for each dataset. Statistical significance of the difference in performance between trials was evaluated using paired t-tests. The results can be seen Figure 2. No bracket indicates no significant difference. The asterisk symbol indicates the right result is significantly better than the left. The triangle symbol indicates that the right is significantly worse. A single symbol indicates the t-test resulted in *p*-value less than 0.05. Two symbols indicate a *p*-value less than 1e-3. Three indicate a *p*-value less than 1e-5.

The level set loss term provided significant improvement for all datasets. The weighted gradient term provided significant improvement for the IAC and Healthy KiTS21 datasets, while making performance on the SegTHOR dataset significantly worse.

For each dataset, the final "best" loss function after dataset-specific loss function customization can be seen in Equations 17-19 below.

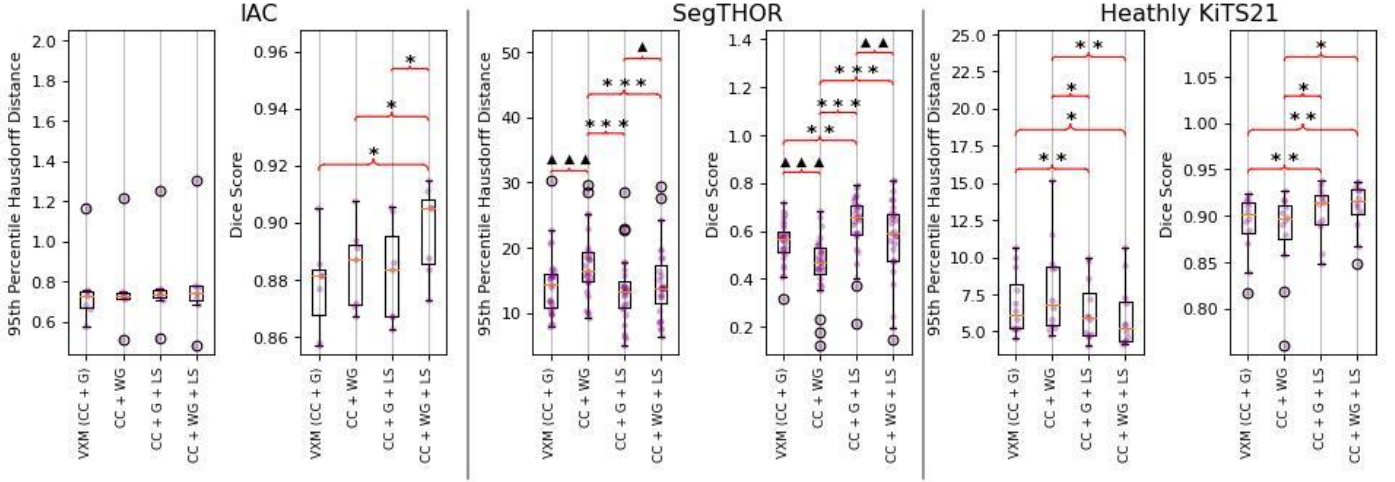

Figure 2. Ablation study results

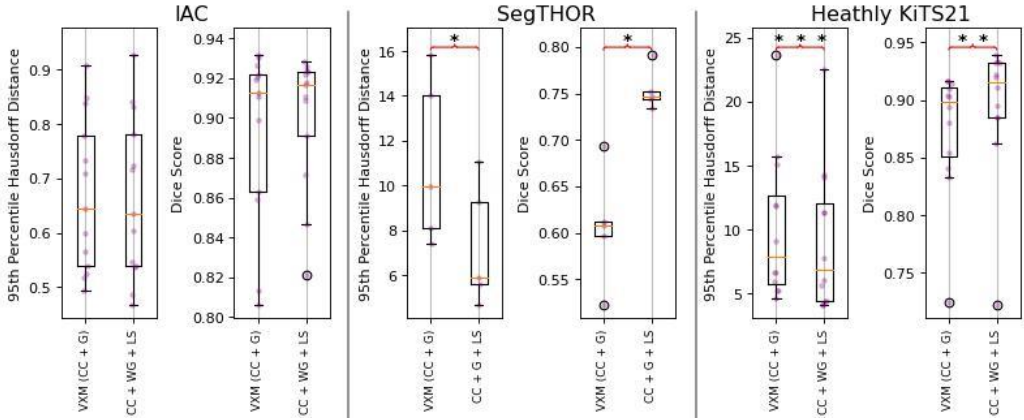

Figure 3. Results of our proposed method versus VoxelMorph VXM. Symbols follow same legend as ablation study.

$$loss_{IAC} = loss_{cc} + loss_{wgrad} + 0.5\, loss_{LS} \quad (17)$$

$$loss_{SegTHOR} = loss_{cc} + loss_{grad} + 0.5\, loss_{LS} \quad (18)$$

$$loss_{HKiTS21} = loss_{cc} + loss_{wgrad} + 0.5\, loss_{LS} \quad (19)$$

Figure 3 shows the results of a network trained with our final loss functions (see Equations 17-19) on the testing datasets. These results are compared to those from a network trained using the loss function used in the VoxelMorph framework (see Equation 14). The significance is displayed in the same manner as in Figure 2.

Qualitative results for each dataset can be seen in Figure 4, Figure 5, and Figure 6, respectively. In each figure, results of the proposed network are shown in blue compared to ground truth in red for the testing case for which the proposed network performed best (left) and worst (right) in terms of Dice score. For the IAC dataset, errors are fairly uniformly distributed and relatively small. For the SegTHOR datasets, the largest errors typically occur at the superior boundary where there is a lack of a true anatomical boundary to indicate the location of the boundary of the structure. Fort the KiTS21 dataset, large errors are observed in some cases, likely due to smaller volume of surrounding darker intensity tissue.

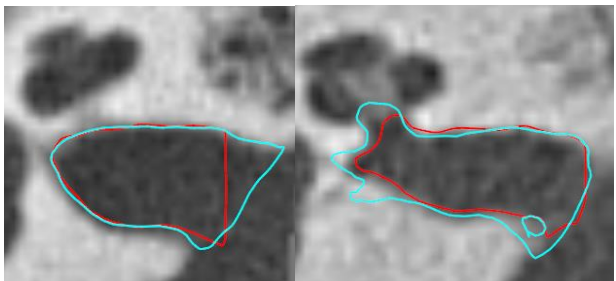

*Figure 4. Axial slices of best (left) and worst (right) cases of our proposed method on IAC testing data. Network and ground truth segmentations are shown with blue and red, respectively.*

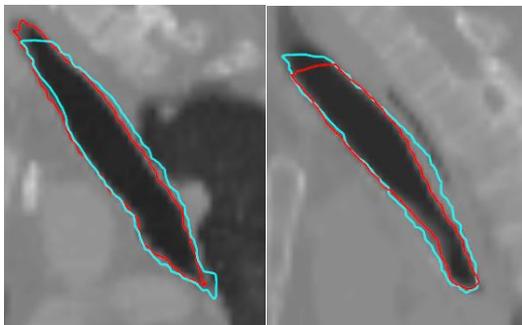

*Figure 5. Coronal slices of best (left) and worst (right) cases of our proposed method on SegTHOR testing data. Network and ground truth segmentations are shown with blue and red, respectively.*

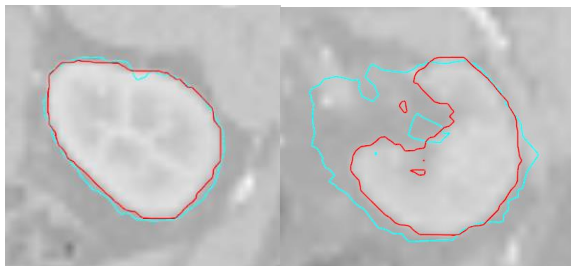

*Figure 6. Axial slices of best (left) and worst (right) cases of our proposed method on HealthyKiTS21 testing data. Network and ground truth segmentations are shown with blue and red, respectively.*

## V. Discussion and Conclusion

The results indicate the significant promise of our proposed extensions to the loss function used in the VoxelMorph framework. All testing accuracy metrics on all datasets show average improvement relative to VoxelMorph. These improvements were statistically significant, with the exception of the $95^{th}$ Percentile Hausdorff distance metric on the IAC dataset, which shows no significant difference.

Our approach has limitations in terms of the types of anatomical structures for which it is applicable. For structures that do not exhibit intensity homogeneity or do not have distinct difference in average intensity values between foreground and background, our proposed loss function will not improve upon the more general loss function used in the VoxelMorph framework. However, we find that for atlas-based segmentation of structures that are reasonably well represented by a Chan and Vese energy functional, our method provides significant improvement.

Future projects include utilizing our proposed method on the IAC dataset to localize auditory nerve fibers. Since ANFs are enclosed by the IAC, we hypothesize that an accurate IAC atlas-based segmentation should facilitate an accurate localization of the ANFs. We hope that the resulting ANF localizations will provide essential information for our patient-specific ANF activation models.

## VI. References

[1] Z. Liu and J. H. Noble, "Auditory Nerve Fiber Health Estimation Using Patient Specific Cochlear Implant Stimulation Models," *Lecture Notes in Computer Science - Proceedings of the International Workshop on,* vol. 12417, pp. 184-184, 2020.

[2] J. H. Noble, R. F. Labadie, R. H. Gifford and B. M. Dawant, "Image-Guidance Enables New Methods for Customizing Cochlear Implant Stimulation Strategies," *IEEE transactions on neural systems and rehabilitation,* pp. 820-829, 2013.

[3] J. Long, E. Shelhamer and T. Darrell, "Fully Convolutional Networks for Semantic Segmentation," arXiv:1411.4038 [cs.CV], 2015.

[4] A. Krizhevsky, I. Sutshever and G. E. Hinton, "Imagenet classification with deep convolutional neural networks," *Advances in Neural Information Processing Systems,* pp. 1106-1114, 2012.

[5] O. Ronneberger, P. Fischer and T. Brox, "U-Net: Convolutional Networks for Biomedical Image Segmentation," in *Medical Image Computing and Computer-Assisted Intervention – MICCAI 2015*, Springer International Publishing, 2015, pp. 234-241.

[6] Ö. Çiçek, A. Abdulkadir, S. S. Lienkamp, T. Brox and O. Ronneberger, "3D U-Net: Learning Dense Volumetric Segmentation from Sparse Annotation," in *Medical Image Computing and Computer-Assisted Intervention MICCAI 2016*, 2016.

[7] A. Wolny, L. Cerrone, A. Vijayan, R. Tofanelli, A. V. Barro, M. Louveaux, C. Wenzl, S. Strauss, D. Wilson-Sánchez, R. Lymbouridou, S. S. Steigleder and others, "Accurate and versatile 3D segmentation of plant tissues at cellular resolution," *eLife,* vol. 9, p. e57613, 2020.

[8] D. Zhang, R. Banalagay, J. Wang, Y. Zhao, J. H. Noble and B. M. Dawant, "Two-level training of a 3D U-Net for accurate segmentation of the intra-cochlear anatomy in head CTs with limited ground truth training

data," *Proceedings of the SPIE Conference on Medical Imaging,* vol. 10949, pp. 1094907-1094907-8, 2019.

[9] H. G. Mason and J. H. Noble, "Automatic internal auditory canal segmentation using a weakly supervised 3D U-Net," *Medical Imaging 2022: Image-Guided Procedures, Robotic Interventions, and Modeling,* vol. 12034, pp. 431-438, 2022.

[10] G. Papandreou, L. Chen, K. Murphy and A. L. Yuille, "Weakly and semi-supervised learning of a deep convolutional network for semantic image segmentation," *2015 IEEE International Conference on Computer Vision (ICCV),* pp. pp. 1742-1750, 2015.

[11] G. Balakrishnan, A. Zhao, M. R. Sabuncu, J. V. Guttag and A. V. Dalca, "VoxelMorph: A Learning Framework for Deformable Medical Image Registration," *IEEE Transactions on Medical Imaging,* vol. 38, pp. 1788-1800, 2019.

[12] N. Miyake, H. Lu, T. Kamiya, T. Aoki and S. Kido, "Temporal Subtraction Technique for Thoracic MDCT Based on Residual VoxelMorph," *Image Processing and Machine Learning in Disease Predictions and Diagnosis,* 2022.

[13] Y. Zhu, Z. Zhou Sr, G. Liao Sr. and K. Yuan, "New loss functions for medical image registration based on VoxelMorph," *Proceedings of the SPIE Conference on Medical Imaging,* vol. 11313, 2020.

[14] D. Mumford and J. Shah, "Optimal Approximations by Piecewise Smooth Functions and Associated Variational Problems," *Communications on Pure and Applied Mathematics,* vol. 42, no. 5, pp. 577-685, 1989.

[15] T. Chan and L. Vese, "Active contours without edges," *IEEE Transactions on Image Processing,* vol. 10, pp. 266-277, 2001.

[16] B. K. a. J. C. Ye, "Mumford-Shah Loss Functional for Image Segmentation With Deep Learning," *IEEE transactions on image processing,* vol. 29, pp. 1856-1866, 2020.

[17] Z. Lambert, C. Petitjean, B. Dubray and S. Ruan, "SegTHOR: Segmentation of Thoracic Organs at Risk in CT images," arXiv:1912.05950 (eess.IV), 2019.

[18] N. Heller, F. Isensee, D. Trofimova, R. Tejpaul, Z. Zhao, H. Chen, L. Wang, A. Golts, D. Khapun, D. Shats, Y. Shoshan, F. Gilboa-Solomon, Y. George, X. Yang, J. Zhang, J. Zhang, Y. Xia, M. Wu, Z. Liu, E. Walczak and S. McSweeney, "The KiTS21 Challenge: Automatic segmentation of kidneys, renal tumors, and renal cysts in corticomedullary-phase CT," arXiv:2307.01984 [cs.CV], 2023.

[19] F. Maes, A. Collignon, D. Vandermeulen, G. Marchal and P. Suetens, "Multimodality image registration by maximization of mutual information," *IEEE Transactions on Medical Imaging,* vol. 16, no. 2, pp. 187-198, 1997.

[20] J. Noble, B. M. Dawant and R. F. L. F. M. Warren, "Automatic Identification and 3D Rendering of Temporal Bone Anatomy," *Otol & Neurotol.,* vol. 30(4), pp. 436-42, 2009.

[21] G. I. Alduraibi, R. Banalagay, R. F. Labadie and J. H. Noble, "Automatic localization of the internal auditory canal for patient-specific cochlear implant modeling," in *Proceedings of SPIE*, San Diego, 2019.

[22] A. J. Al-Shareef, Abbod and M. F., "Neural Networks Initial Weights Optimisation," *2010 12th International Conference on Computer Modelling and Simulation,* pp. 57-61, 2010.

[23] Y.-T. Chang, J. Lin and J.-S. Shieh, "Optimization the initial weights of artificial neural networks via genetic algorithm applied to hip bone fracture prediction," *Advances in Fuzzy Systems,* vol. 2012, no. 6, p. 6, 2012.

[24] "Reproducibility," PyTorch, [Online]. Available: https://pytorch.org/docs/stable/notes/randomness.html. [Accessed 30 10 2023].

[25] P. Henderson, R. Islam, P. Backman, J. Pineau, D. Precup and D. Meger, "Deep Reinforcement Learning that Matters," *Thirthy-Second AAAI Conference On Artificial Intelligence (AAAI),* vol. 1, 2018.

[26] H. G. Mason and J. H. Noble, "Atlas-based automatic internal auditory canal localization with a weakly-supervised 3D U-Net," *Image-Guided Procedures, Robotic Interventions, and Modeling,* no. SPIE, pp. 431-438, 2023.